\documentclass{elsart}
\usepackage{epsfig}
\usepackage{amssymb}

\newcommand{\sink}{\mathop{\rm sink}}
\newcommand{\R}{{\mathbf{R}}}
\renewcommand{\P}{\mathrm{P}}
\newcommand{\V}{{\mathcal V}}
\newcommand{\E}{{\mathcal E}}
\newcommand{\G}{{\mathcal G}}
\newcommand{\keyw}[1]{\textnormal{\textrm{\texttt{#1}}}}
\newcommand{\vi}{{\sf V}}
\newcommand{\bfaceq}{{{\mathcal B}/\!\sim}}
\newcommand{\figeps}[3]{
  \begin{figure}
  \begin{center}
  \epsfig{figure=#1.eps, width=#2cm}
  \end{center}
  \caption{#3}
  \label{fig:#1}
  \end{figure}
}

\begin{document}
\begin{frontmatter}

\title{Violator Spaces: Structure and Algorithms
\thanksref{label1}}
\thanks[label1]{The first and the third author acknowledge support 
from the Swiss Science Foundation (SNF), Project No.\ 200021-100316/1.
The fourth author acknowledges support from the Czech Science Foundation
(GACR), Grant No.\ 201/05/H014.}
\author[label2]{B. G\"artner},
\ead{gaertner@inf.ethz.ch}
\author[label3]{J. Matou\v{s}ek},
\ead{matousek@kam.mff.cuni.cz}
\author[label2]{L. R\"ust},
\ead{ruestle@inf.ethz.ch}
\author[label3]{P. \v{S}kovro\v{n}}
\ead{xofon@kam.mff.cuni.cz}
\address[label2]{
Institute of Theoretical Computer Science,
ETH Z\"urich,
8092 Z\"urich, Switzerland 
}
\address[label3]{
Department of Applied Mathematics and Institute of Theoretical Computer Science,
Charles University,
Malostransk\'{e} n\'{a}m. 25, 
118~00~~Praha~1, Czech Republic
}

\begin{abstract}
Sharir and Welzl introduced an abstract framework for optimization 
problems, called {\em LP-type problems\/} or also {\em generalized 
linear programming problems}, which proved useful in algorithm design.
We define a new, and as we believe, simpler and more natural framework: 
{\em violator spaces}, which constitute a proper generalization of LP-type 
problems. We show that Clarkson's randomized algorithms for low-dimensional 
linear programming work in the context of violator spaces. For example, 
in this way we obtain the fastest known algorithm for the  \emph{$\P$-matrix 
generalized linear complementarity problem} with a constant number of blocks. 
We also give two new characterizations of LP-type problems: they are 
equivalent to {\em acyclic\/} violator spaces, as well as to {\em concrete\/}
LP-type problems (informally, the constraints in a concrete LP-type
problem are subsets of a linearly ordered ground set, and the value
of a set of constraints is the minimum of its intersection).
\end{abstract}

\begin{keyword}
  LP-type problem \sep generalized linear programming \sep violator
  space \sep Clarkson's algorithms \sep unique sink orientation \sep
  generalized linear complementarity problem
\end{keyword}

\end{frontmatter}

\section{Introduction}
The framework of LP-type problems,
invented by Sharir and Welzl in 1992 \cite{sw-cblpr-92},
has become a well-established tool
in the field of geometric optimization. Its origins are in linear
programming: Sharir and Welzl developed a randomized variant of the 
dual simplex algorithm for linear programming and showed that this
algorithm actually works for a more general class of problems they
called LP-type problems. 

For the theory of linear programming, this
algorithm constituted an important progress, since it was later shown 
to be \emph{subexponential} in the RAM model \cite{msw-sblp-92}. 
Together with a similar result  independently obtained by Kalai
\cite{k-srsa-92}, this was the first linear programming algorithm
provably requiring a number of arithmetic operations subexponential
in the dimension and number of constraints (independent of the
precision of the input numbers).

For many other geometric optimization problems in fixed dimension, 
the algorithm by Sharir and Welzl was the first to achieve expected
linear runtime, simply because these problems could be formulated
as LP-type problems. The class of LP-type problems for example
includes the problem of computing the minimum-volume ball or
ellipsoid enclosing a given point set in $\R^d$, and the problem
of finding the distance of two convex polytopes in $\R^d$. Many
other problems have been identified as LP-type problems over the
years \cite{msw-sblp-92,a-bbhdn-94,a-httgl-94,bsv-dsapg-01,halman}.

Once it is shown that a particular optimization problem is an LP-type
problem, and certain algorithmic primitives are implemented for it,
several efficient algorithms are immediately at our disposal: the
Sharir--Welzl algorithm, two other randomized optimization algorithms due
to Clarkson \cite{c-lvali-95} (see \cite{gw-lpraf-96,cm-ltdao-96} for
a discussion of how it fits the LP-type framework), a deterministic
version of it \cite{cm-ltdao-96}, an algorithm for computing the
minimum solution that violates at most $k$ of the given $n$
constraints \cite{m-gofvc-95}, and probably more are to come in 
the future.

The framework of LP-type problems is not only a prototype for concrete
optimization problems, it also serves as a mathematical tool by
itself, in algorithmic \cite{GWSampl01,ChanTukey} and non-algorithmic
contexts \cite{a-spiht-96}. 

An (abstract) LP-type problem is given by a finite set $H$ of 
{\em constraints\/} and a {\em value\/} $w(G)$ for every  subset 
$G\subseteq H$. The values can be real numbers or, for technical convenience, 
elements of any other linearly ordered set. Intuitively, $w(G)$ is 
the minimum value of a solution that satisfies all constraints in $G$. 
The assignment $G\mapsto w(G)$ has to obey the axioms  in the 
following definition.

\begin{defn}\label{def:lptype}
An {\em abstract LP-type problem} is a quadruple $(H,w,W,\leq)$, where
$H$ is a finite set, $W$ is a set linearly ordered by $\leq$, and
$w\colon 2^H\to W$ is a mapping satisfying the following two conditions:

\begin{tabular}{ll}
Monotonicity: & for all $F\subseteq G\subseteq H$ we have 
$w(F)\leq w(G)$, and\\
Locality: & for all $F\subseteq G\subseteq H$ and all $h\in H$ with 
$w(F)=w(G)$ and\\
&$w(G)<w(G\cup\{h\})$, we have $w(F)<w(F\cup\{h\})$.\\
\end{tabular}
\end{defn}

As our running example, we will use the smallest enclosing ball problem, 
where $H$ is a finite point set in $\R^d$ and $w(G)$ is the radius of 
the smallest ball that encloses all points of $G$. 
In this case monotonicity is obvious, while verifying locality requires 
the nontrivial but well known geometric result that the smallest 
enclosing ball is unique for every set.

It seems that the order $\leq$ of subsets is crucial; after all,
LP-type problems model \emph{optimization problems}, and indeed,
the subexponential algorithm for linear programming and other 
LP-type problems \cite{msw-sblp-92} heavily relies on such an 
order. 

A somewhat deeper look reveals that often, we only care whether two
subsets have the \emph{same} value, but not how they compare under
the order $\leq$. The following definition is taken from
\cite{sw-cblpr-92}:

\begin{defn}\label{def:lp_type_basis}
Consider an abstract LP-type problem $(H,w,W,\leq)$. We say that 
$B\subseteq H$ is a {\em basis} if for all proper subsets $F\subset B$ we 
have $w(F)\neq w(B)$. For $G\subseteq H$, a \emph{basis of $G$} is a minimal 
subset $B$ of $G$ with $w(B)=w(G)$.
\end{defn}

We observe that a minimal subset $B\subseteq G$ with $w(B)=w(G)$ is
indeed a basis.

Solving an abstract LP-type problem $(H,w,W,\leq)$ means to find a
basis of $H$. In the smallest enclosing ball problem, a basis of $H$
is a minimal set $B$ of points such that the smallest enclosing ball
of $B$ has the same radius (and is in fact the same) as the smallest
enclosing ball of $H$, $w(B)=w(H)$.

In defining bases, and in saying what it means to solve an LP-type
problem, we therefore do not need the order $\leq$. The 
main contribution of this paper is that many of the things one can
prove about LP-type problems do not require a concept of order.

We formalize this by defining the new framework of 
\emph{violator spaces}. 
Intuitively, a violator space is an LP-type problem without order.
This generalization of LP-type problems is proper, and we can exactly
characterize the violator spaces that ``are'' LP-type problems. 
In doing so, we also establish yet
another equivalent characterization of
LP-type problems that is closer to the applications than the abstract
formulation of Definition \ref{def:lptype}. In a concrete LP-type
problem, the constraints are not just elements of a set, but 
they are associated with subsets of some linearly ordered ground 
set $X$, with the minimal elements in the intersections of such 
subsets corresponding to ``solutions''. 
The framework of concrete LP-type problems is similar to the model
presented in \cite{a-httgl-94} as a mathematical programming problem, with
a few technical differences.

These are our main findings on the structural side. Probably the most 
surprising insight on the algorithmic side is that 
Clarkson's algorithms \cite{c-lvali-95} work for violator spaces
of fixed dimension, leading to an expected linear-time algorithm
for ``solving'' the violator space.
Clarkson's algorithms were originally developed for linear
programs with small dimension. They can be
generalized for LP-type problems \cite{gw-lpraf-96,cm-ltdao-96}. The
fact that the scheme also works for violator spaces may come as a 
surprise since the structure of
violator spaces is not acyclic in general (in contrast to LP-type
problems). The LP-type algorithm from \cite{msw-sblp-92} is also
applicable to violator spaces, but its analysis breaks down.

We give an application of Clarkson's algorithms in the more general
setting by linking our new violator space
framework to well-known abstract and concrete frameworks in
combinatorial optimization. For this, we show that any \emph{unique
sink orientation} (USO) of the cube 
\cite{SW,grid_uso,M02,MatUSO,MCube,Develin,MS,SchSz,SS,lptouso} 
or the more general grid \cite{grid_uso} gives rise to a violator 
space, but not to an LP-type problem in general. 
Grid USO capture some important problems
like linear programming over products of simplices, \emph{generalized
linear complementarity problems} over $\mathrm{P}$-matrices \cite{grid_uso}
or games like parity, mean-payoff, and simple stochastic games
\cite{sweden1,sweden2,ssg_pglcp}.

We show that we can find the sink in a unique sink orientation by
solving the violator space, for example with Clarkson's
algorithms. A concrete new result is obtained by applying this to
$\P$-matrix generalized linear complementarity problems. 
These problems are not known to be polynomial-time solvable, but
NP-hardness would imply NP=co-NP \cite{Meg,grid_uso}. 
Since any $\P$-matrix generalized linear
complementarity problem gives rise to a unique sink orientation
\cite{grid_uso}, we may use violator spaces and Clarkson's algorithms
to solve the problem in expected linear time in the (polynomially
solvable) case of a \emph{fixed} number of \emph{blocks}. This is
optimal and beats all previous algorithms.

The rest of the paper is organized as follows. In Section
\ref{sec:basics}, we formally define the frameworks of
concrete LP-type problems and violator spaces, along with their
essential terminology. Then we state our main structural result.

In Section \ref{sec:mainproof}, we prove this result by deriving the
equivalence of abstract and concrete LP-type problems, and of
\emph{acyclic} violator spaces. 

Section \ref{sec:clarkson} shows that Clarkson's algorithms work for
(possibly cyclic) violator spaces. Section \ref{grid_uso}, finally,
shows how unique sink orientations induce violator spaces. A
unique sink orientation can be cyclic, and a cyclic orientation gives
rise to a cyclic violator space. Unique sink orientations are
therefore nontrivial examples of possibly cyclic violator spaces.

\section{Structural Results}\label{sec:basics}

\subsection{Concrete LP-type problems.}
Although intuitively one thinks about $w(G)$ as the value of an
optimal solution of an optimization problem, the solution itself is
not explicitly represented in Definition \ref{def:lptype}.  In
specific geometric examples, the constraints can usually be
interpreted as a subset of some ground set $X$ of points, and the
optimal solution for $G$ is the point with the smallest value in the
intersection of all constraints in $G$. For example, in linear
programming, the constraints are halfspaces, the value is given by the
objective function, and the optimum is the point with minimum value in
the admissible region, i.e., the intersection of the halfspaces. In
order to have a unique optimum for every set of constraints (which is
needed for $w$ to define an LP-type problem), one assumes that the
points are linearly ordered by the value; for linear programming, we
can always take the lexicographically smallest optimal solution, for
instance.

Such an interpretation is possible for the smallest enclosing ball problem 
too, although it looks a bit artificial. Namely, the ``points'' of $X$ 
are all balls in $\R^d$, where the ordering can be an arbitrary linear 
extension of the partial ordering of balls by radius. The ``constraint''
for a point $h\in H$ is the set of all balls containing $h$.

The following definition captures this approach to LP-type problems.

\begin{defn}
\label{def:concrete_LPtype}
A {\em concrete LP-type problem} is a triple $(X,\preceq,{\mathcal H})$, 
where $X$ is a set linearly ordered by $\preceq$, 
${\mathcal H}$ is a finite multiset whose elements are subsets of
$X$, and for any 
${\mathcal G}\subseteq{\mathcal H}$, if the intersection 
$\bigcap{\mathcal G}:=\bigcap_{G\in\mathcal{G}}G$ is nonempty, then it 
has a minimum element with respect to $\preceq$ (for ${\mathcal G}=\emptyset$ 
we define $\bigcap{\mathcal G}:=X$).
\end{defn}

The definition allows $\mathcal H$ to be a multiset, i.e., a constraint
set $A\subseteq X$ may be included several times.
For example, in an instance of linear programming, some constraints
can be the same, which we can reflect by this.
In Subsection \ref{sec:examples} we provide an example of an abstract 
LP-type problem, for which the multiplicity comes in handy to represent it as
a concrete LP-type problem.

A similar model has been presented in \cite{a-httgl-94} (mathematical
programming problem). The slight difference is that it allows several
points to have the same value but the constraints form a set rather than
a multiset.

Bases are defined analogously to Definition \ref{def:lp_type_basis}.

\begin{defn}\label{def:concrete_basis}
Consider a concrete LP-type problem $(X,\preceq,{\mathcal H})$. We say that 
$\mathcal{B}\subseteq\mathcal{H}$ is a {\em basis} if for all proper submultisets 
$\mathcal{F}\subset\mathcal{B}$ we have $\min(\bigcap\mathcal{F})\prec
\min(\bigcap\mathcal{B})$. For $\mathcal{G}\subseteq\mathcal{H}$, a 
\emph{basis of $\mathcal{G}$} is a minimal $\mathcal{B}\subseteq\mathcal{G}$
with $\min(\bigcap\mathcal{B})=\min(\bigcap\mathcal{G})$.
\end{defn}

As before, a minimal $\mathcal{B}\subseteq\mathcal{G}$ with 
$\min(\bigcap\mathcal{B})=\min(\bigcap\mathcal{G})$ is indeed a basis. 

Given any concrete LP-type problem ${\mathcal P}=(X,\preceq,{\mathcal H})$,
we obtain an abstract LP-type problem $P=({\mathcal H},w,X,\preceq)$ 
according to Definition \ref{def:lptype} by putting 
$w({\mathcal G})=\min(\bigcap {\mathcal G})$ (or $w({\mathcal G})=+\infty$, 
if $\bigcap{\mathcal G}$ is empty), as is easy to check (proof omitted). 
It is clear that $\mathcal{B}\subseteq\mathcal{G}$ is a basis of 
$\mathcal{G}$ in $P$ if and only if $\mathcal{B}$ is a basis of 
$\mathcal{G}$ in $\mathcal{P}$. We say that $P$ is 
\emph{basis-equivalent} to $\mathcal{P}$.

Somewhat surprising is the converse, which we prove below in 
Theorem~\ref{thm:main}:
Any abstract LP-type problem $(H,w,W,\leq)$ has a 
``concrete representation'', that is, a concrete LP-type problem 
that is basis-equivalent to $(H,w,W,\leq)$.

Strictly speaking, if the multiset ${\mathcal H}$ in the 
concrete LP-type problem has elements with
multiplicity bigger than 1, then ${\mathcal H}$ cannot be used
as the set of constraints for the abstract LP-type problem (since it
is not a set).
However, we can bijectively map $\mathcal H$ to a set, i.e., we take
any set $H$ with $|H|=|{\mathcal H}|$ and a mapping
$f\colon H\to{\mathcal H}$ such that for any $\bar h\in{\mathcal H}$,
the number of elements $h\in H$ that map to $\bar h$ is equal to the
multiplicity of $\bar h$. For $G\subseteq H$ we then define
$w(G)=\min(\bigcap_{g\in G} f(g))$ which gives us a fair abstract LP-type
problem $P=(H,w,X,\preceq)$ basis-equivalent to $\mathcal{P}$.
In this case, by basis-equivalence we mean the existence of a 
suitable mapping $f$ together with the condition that $B\subseteq G$
is a basis of $G$ in $P$ if and only if the multiset 
$\{f(b)\colon b\in B\}$ is a basis of $\{f(g)\colon g\in G\}$ in
$\mathcal{P}$.

\subsection{Violator spaces.} Let $(H,w,W,\leq)$ be an abstract 
LP-type problem. It is natural to define that a constraint $h\in H$ 
{\em violates\/} a set $G\subseteq H$ of constraints if 
$w(G\cup\{h\})>w(G)$. For example, in the smallest enclosing ball problem, 
a point $h$ violates a set $G$ if it lies outside of the smallest ball 
enclosing $G$ (which is unique).

\begin{defn}
The \emph{violator mapping} of $(H,w,W,\leq)$ is defined by
$\vi(G)=\{h\in H\colon w(G\cup\{h\})>w(G)\}$. Thus, $\vi(G)$ is the set
of all constraints violating $G$. 
\end{defn}

It turns out that the knowledge of
$\vi(G)$ for all $G\subseteq H$ is enough to describe the ``structure'' 
of an LP-type problem. That is, while we cannot reconstruct $W$, $\leq$, 
and $w$ from this knowledge, it is natural to consider two LP-type problems 
with the same mapping $\vi\colon 2^H\to 2^H$ the same (isomorphic). Indeed, 
the algorithmic primitives needed for implementing the Sharir--Welzl 
algorithm and the other algorithms for LP-type problems mentioned above 
can be phrased in terms of testing violation (does $h\in \vi(G)$ hold for 
a certain set $G\subseteq H$?), and they never deal explicitly with the 
values of $w$.

We now introduce the notion of {\em violator space}:

\begin{defn}\label{def:vs}
A {\em violator space} is a pair $(H,\vi)$, where $H$ is a finite set 
and $\vi$ is a mapping $2^H\to2^H$ such that 

\begin{tabular}{ll}
Consistency: & $G\cap \vi(G)=\emptyset$ holds for all $G\subseteq H$, and\\
Locality: & for all $F\subseteq G\subseteq H$, where 
$G\cap \vi(F)=\emptyset$, we have\\
& $\vi(G)=\vi(F)$.\\
\end{tabular}
\end{defn}

A basis of a violator space is defined in analogy to a basis of an
LP-type problem.

\begin{defn}\label{def:vbasis}
Consider a violator space $(H,\vi)$. We say that $B\subseteq H$ 
is a {\em basis} if for all proper subsets $F\subset B$ we have 
$B\cap \vi(F)\neq\emptyset$. For $G\subseteq H$, a \emph{basis of $G$} 
is a minimal subset $B$ of $G$ with $\vi(B)=\vi(G)$.
\end{defn}

Observe that a minimal subset $B\subseteq G$ with $\vi(B)=\vi(G)$ is 
indeed a basis: Assume for contradiction that there is a set
$F\subset B$ such that $B\cap \vi(F)=\emptyset$. Locality then
yields $\vi(B)=\vi(F)=\vi(G)$, which contradicts minimality of $B$.

We will check in Subsection~\ref{sec:abstract->acyclicVS}
that the violator mapping of an abstract LP-type
problem satisfies the two axioms above. Consistency is immediate:
since $w(G)=w(G\cup\{h\})$ for $h\in G$, no element in $G$ violates 
$G$. The locality condition has the following intuitive interpretation: 
adding only non-violators to a set does not change the value.

We actually show more: given an abstract LP-type problem 
$(H,w,W,{\leq})$, the pair $(H,\vi)$, with $\vi$ being the violator mapping, 
is an \emph{acyclic} violator space. (Acyclicity of a violator space
will be defined later in Definition \ref{def:vsetord}.) It turns out 
in Subsection \ref{sec:acyclicVS->concrete} 
that acyclicity already characterizes the 
violator spaces obtained from LP-type problems, and thus any acyclic 
violator space can be represented as an LP-type problem (abstract or 
concrete). These equivalences are stated in our main theorem.

\begin{thm}\label{thm:main}
The axioms of abstract LP-type problems, of concrete LP-type problems, 
and of acyclic violator spaces are equivalent. More precisely, every
problem in one of the three classes has a basis-equivalent problem
in each of the other two classes.
\end{thm}

The construction is illustrated on simple instances of problems of
linear programming and the smallest enclosing ball in
Subsection \ref{sec:examples}. Several more results 
concerning violator spaces have been achieved in the MSc. thesis of 
the fourth author \cite{Skovron}.

\section{Equivalence of LP-type Problems and Acyclic Violator Spaces}
\label{sec:mainproof}
In this section we prove Theorem \ref{thm:main}. 

\subsection{Preliminaries on Violator Spaces}\label{sec:preliminaries}

To show that every acyclic violator space $(H,\vi)$ originates from 
some concrete LP-type problem, we need an appropriate linearly 
ordered set $X$ of ``points'', and then we will identify the elements of 
$H$ with certain subsets of $X$.

What set $X$ will we take? Recall that for smallest enclosing balls,
$X$ is the set of all balls, and the subset for $h\in H$ is the subset
of balls containing $h$. It is not hard to see that we may restrict
$X$ to smallest enclosing balls of \emph{bases}; in fact, we may
choose $X$ as the set of bases, in which case the subset for $h$
becomes the set of bases not violated by $h$.

This also works for general acyclic violator spaces, with bases suitably
ordered. The only blemish is that we may get several minimal bases for
$G\subseteq H$; for smallest enclosing balls, this corresponds to the
situation in which several bases define the same smallest enclosing
ball. To address this, we will declare such bases as equivalent
and choose $X$ as the set of all equivalence classes instead.

In the following, we fix a violator space $(H,\vi)$. The set of all 
bases in $(H,\vi)$ will be denoted by $\mathcal B$.

\begin{defn}\label{def:vequiv}
$B,C\in \mathcal B$ are \emph{equivalent}, $B\sim C$, if $\vi(B)=\vi(C)$.
\end{defn}

Clearly, the relation $\sim$ defined on $\mathcal{B}$ is an equivalence
relation. The equivalence class containing a basis $B$ will be
denoted by $[B]$.

Now we are going to define an ordering of the bases, and we derive from this
an ordering of the equivalence classes as well as the notion of acyclicity 
in violator spaces.

\begin{defn}\label{def:vsetord}
For $F,G\subseteq H$ in a violator space $(H,\vi)$, we say that 
$F\leq_0 G$ ($F$ is {\em locally smaller} than $G$) if 
$F\cap \vi(G)=\emptyset$.

For equivalence classes $[B],[C]\in\bfaceq$, we say that 
$[B]\leq_0 [C]$ if there exist  $B'\in[B]$ and $C'\in[C]$ such 
that $B'\leq_0 C'$.

We define the relation $\leq_1$ on the equivalence classes as 
the transitive closure of $\leq_0$. The relation $\leq_1$ 
is clearly reflexive and transitive. If it is antisymmetric, we say that the 
violator space is {\em acyclic}, and we define the relation $\leq$ 
as an arbitrary linear extension of $\leq_1$.
\end{defn}

The intuition of the \emph{locally-smaller} notion comes from LP-type
problems: if no element of $F$ violates $G$, then $G\cup F$ has the
same value as $G$ (this is formally proved in Lemma
\ref{lem:conseqloc} below), and monotonicity yields that value-wise,
$F$ is smaller than or equal to $G$.

Note that in the definition of $[B]\leq_0[C]$ we do not require 
$B'\leq_0 C'$ to hold for {\em every} $B'$ and $C'$. In fact
$B'\not\leq_0 C'$ may happen for some bases $B'$ and $C'$, 
but $C'\leq_0 B'$ can not hold (which can easily be shown). 
%
%Assume for contradiction
%that $[B]\leq_0 [C]$ and $[C]\leq_0 [B]$. Then $B'\cap V(C')=\emptyset$
%and $C'\cap V(B')=\emptyset$ for some $B'$ and $C'$ out of the
%respective equivalence classes (because all bases in the same
%equivalence class have the same violators). This is the same as
%$(B'\cup C')\cap V(C')=\emptyset$ and $(B'\cup C')\cap V(B')=\emptyset$,
%which with locality yields the contradiction $V(C')=V(B'\cup C')=V(B')$.

To show that acyclicity does not always hold, we conclude this
section with an example of a cyclic violator space.

We begin with an intuitive geometric description; see Figure \ref{fig:cyclic}. 
We consider a triangle without the center point. We say that a point is 
``locally smaller'' if it is farther clockwise with respect to the center. 
The constraints in our violator space are the three halfplanes $f,g,h$.

\figeps{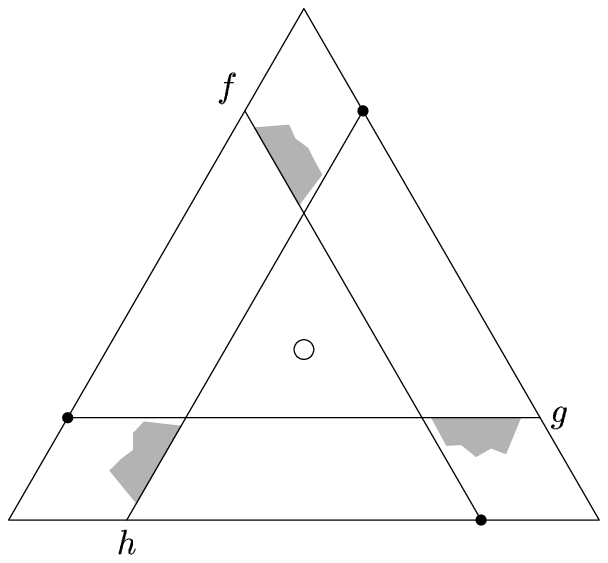}{5}{A cyclic violator space.}

The locally smallest point within each halfplane is marked, and a halfplane
violates a set of halfplanes if it does not contain the locally smallest point
in their intersection.

Now we specify the corresponding violator space formally.
We have $H=\{f,g,h\}$, and $\vi$ is given by the following table:
\begin{center}
\tabcolsep0.3cm
\begin{tabular}{|c||c|c|c|c|c|c|c|c|}\hline
$G$ & $\emptyset$ & $f$ & $g$ & $h$ & $f,g$ & $f,h$ & $g,h$ & $f,g,h$ \\ \hline
$\vi(G)$ & $f,g,h$ & $h$ & $f$ & $g$ & $h$ & $g$ & $f$ & $\emptyset$ \\ \hline
\end{tabular}
\end{center}
This $(H,\vi)$ is really a violator space, since we can easily check both
consistency and locality. The bases are $\emptyset$, one-element sets, and
$H$. We have $\{f\}\leq_0\{h\}\leq_0\{g\}\leq_0\{f\}$, but none of 
the one-element bases are equivalent; i.e., $\leq_1$ is not antisymmetric.

\subsection{Abstract LP-type Problems yield Acyclic Violator Spaces}
\label{sec:abstract->acyclicVS}
In this subsection, we show that the violator mapping of 
an abstract LP-type problem is an acyclic violator space. 
To this end, we need the following two lemmas.

\begin{lem}\label{lem:conseqloc}
Consider an abstract LP-type problem $(H,w,W,\leq)$ with violator mapping 
$\vi$. Let $A,B\subseteq H$, where $B$ is not violated by any $h\in A$ 
($A\cap \vi(B)=\emptyset$). Then $w(A\cup B)=w(B)$.
\end{lem}

\pf From monotonicity, we immediately obtain the inequality ``$\geq$''.
The inequality ``$\leq$'' can be shown by induction on $|A|$.
If $|A|=1$, i.e., $A=\{h\}$, then $w(B\cup\{h\})>w(B)$ would imply that
$B$ is violated by $h\in A$, a contradiction.

Let $|A|>1$ and $A=A_0\mathbin{\dot\cup}\{h\}$ (disjoint union). 
From the induction hypothesis we have $w(B\cup A_0)=w(B)$. Now, if 
$w(B\cup A_0)<w(B\cup A_0\cup \{h\})$, then by locality (for $B\cup A_0$, 
$B$ and $h$) we get $w(B)<w(B\cup\{h\})$. This means that $h\in\vi(B)$,
and since $h\in A$ we have $h\in A\cap\vi(B)$, a contradiction. 
So $w(B)=w(B\cup A_0) \geq w(B\cup A_0\cup\{h\})=w(B\cup A)$. We have
proved $w(A\cup B)\leq w(B)$.
\qed

\begin{lem}\label{lem:separbasis}
Consider an abstract LP-type problem $(H,w,W,\leq)$ with violator mapping 
$\vi$. Then for any $A,B\subseteq H$ with $\vi(A)=\vi(B)$ we have $w(A)=w(B)$. 
Conversely, $w(A)=w(B)=w(A\cup B)$ implies $\vi(A)=\vi(B)$. In particular, 
if $A\subseteq B$ and $w(A)=w(B)$, then $\vi(A)=\vi(B)$.
\end{lem}

Note that the condition $w(A)=w(B)$ generally does not suffice for
$\vi(A)=\vi(B)$. For example, having any $H$, we can define $w$ by $w(G)=|G|$ 
for all $G\subseteq H$ (it can be checked that it is an abstract LP-type
problem). Then any $G$'s of the same size 
have the same $w$, however, $\vi(G)=H\setminus G$, and so no distinct 
$G$'s share the value of $\vi$. Roughly speaking, the equality 
$w(A)=w(B)$ may hold just ``by accident''. This is one way in which 
we can see that $w$ by itself does not reflect the combinatorial structure
of the problem in a natural way.

\begin{pf*}{PROOF of Lemma \ref{lem:separbasis}.} Let $w(A)\neq w(B)$. 
Without loss of generality we assume $w(A)>w(B)$ (note that here we use 
the linearity of the ordering $\leq$). If $A\cap \vi(B)=\emptyset$, from 
Lemma \ref{lem:conseqloc} we would get $w(A\cup B)=w(B)$, which contradicts 
$w(A\cup B)\geq w(A)>w(B)$. So there necessarily exists $h\in A\cap \vi(B)$, 
but since $h\in A$, we have $h\not\in \vi(A)$. So $\vi(A)\neq \vi(B)$.

Conversely, suppose $w(A)=w(B)=w(A\cup B)$. We want to show $\vi(A)=\vi(B)$, 
i.e., that $w(A)<w(A\cup\{h\})$ holds iff  $w(B)<w(B\cup\{h\})$
holds. By symmetry, it suffices to show only one of the implications.
We assume $w(A)<w(A\cup\{h\})$. Then $w(A\cup B)=w(A)<w(A\cup\{h\}) \leq
w(A\cup B\cup\{h\})$. Since $B\subseteq A\cup B$ and $w(B)=w(A\cup B)$, we
may use locality, which gives $w(B)<w(B\cup\{h\})$. So the desired
equivalence holds.
\qed
\end{pf*}

\begin{prop}\label{prop:abstract->acyclicVS}
Consider an abstract LP-type problem $(H,w,W,\leq)$, and let $\vi$ be its
violator mapping. Then $(H,\vi)$ is an acyclic violator space.
Moreover, $(H,\vi)$ is basis-equivalent to $(H,w,W,\leq)$.
\end{prop}

\pf Clearly $G\cap\vi(G)=\emptyset$, since $w(G\cup\{g\})=w(G)$  
for any $g\in G$, so consistency holds. If $G\cap \vi(F)=\emptyset$
for $F \subseteq G$, then by Lemma \ref{lem:conseqloc} we get 
$w(F\cup G)=w(F)$. Since $F\subseteq G$, we have $F \cup G=G$ and so 
$w(G)=w(F)$. Lemma \ref{lem:separbasis} then yields $\vi(G)=\vi(F)$, 
so locality holds.

We proceed to prove acyclicity of $(H,\vi)$. Fix $[B]$ and $[C]$,
$[B]\neq [C]$, with $[B]\leq_0[C]$, that is $B'\cap \vi(C')=\emptyset$ for
some $B'\in [B]$ and $C'\in [C]$. Lemma \ref{lem:conseqloc} implies
$w(C')=w(B'\cup C')$. For contradiction, assume $w(B')\geq w(C')$; then 
$w(B')\geq w(B'\cup C')$ which with monotonicity yields $w(B')=w(B'\cup C')=w(C')$. 
Lemma \ref{lem:separbasis} gives $\vi(B')=\vi(C')$, a contradiction to 
$[B]\neq [C]$. 
Thus $[B]\leq_0[C]$ for $[B]\neq [C]$ implies $w(B')<w(C')$ for some 
bases $B'$ and $C'$ out of the respective equivalence classes. By 
Lemma \ref{lem:separbasis}, $w(B')$ is the same for all $B'\in [B]$ 
(because all bases in $[B]$ have the same violators). Therefore, by chaining 
several $\leq_0$'s we also get $w(B')<w(C')$ for $[B]\leq_1[C]$. This
proves that $\leq_1$ is necessarily antisymmetric (since $\leq$ is an 
ordering of $W$).

Finally, observe that by Lemma \ref{lem:separbasis}, $B\subseteq G$ is an 
inclusion-minimal subset of $G$ with $w(B)=w(G)$ if and only if $B$ is 
an inclusion-minimal subset of $G$ with $\vi(B)=\vi(G)$. So, $B\subseteq G$ 
is a basis of $G$ in $(H,w,W,\leq)$ if and only if $B$ is a basis of $G$ 
in $(H,\vi)$. Thus $(H,\vi)$ is basis-equivalent to $(H,w,W,\leq)$.
\qed

At first glance, one might think that for $F\subseteq G$ we should
have $\vi(F)\supseteq \vi(G)$. Unfortunately, this is not the case, as
the  linear programming example in Figure \ref{fig:counterexample} shows (the 
$y$-coordinate is to be minimized).

\figeps{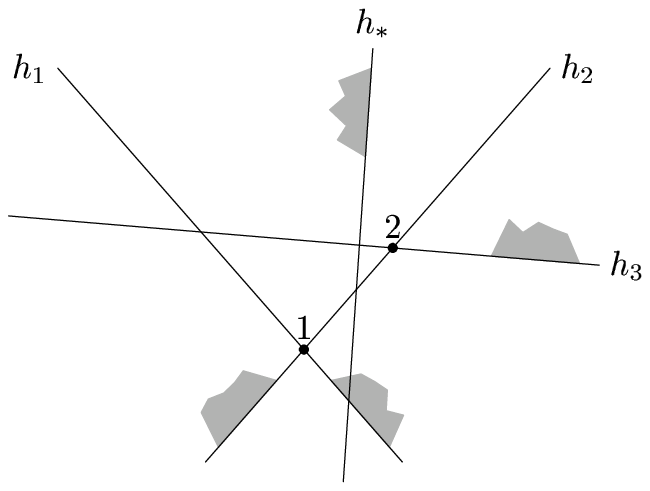}{6}{A linear programming example ($F=\{h_1,h_2\}
\subseteq G=\{h_1,h_2,h_3\}$ with $\vi(G)\not\subseteq\vi(F)$).}

We put $F=\{h_1,h_2\}$ and $G=\{h_1,h_2,h_3\}\supseteq F$. The point $1$
is minimum in the intersection of $F$, and $2$ is minimum in the
intersection of $G$. We have $1\in h^*$, $2\not\in h^*$, and so
$h^*\not\in\vi(F)$ and $h^*\in\vi(G)$.

\subsection{Acyclic Violator Spaces yield Concrete LP-type Problems}
\label{sec:acyclicVS->concrete}
The following proposition is the last ingredient for Theorem \ref{thm:main}.
\begin{prop}\label{prop:acyclicVS->concrete}
Every acyclic violator space $(H,\vi)$ can be represented as a concrete 
LP-type problem that is basis-equivalent to $(H,\vi)$.
\end{prop}

\pf We are given an acyclic violator space $(H, \vi)$ and we define the 
mapping $S\colon H\to 2^\bfaceq$ that will act as a ``concretization'' 
of the constraints in $H$:
\[
S(h)=\{[B]\colon B\in{\mathcal B},\, h\not\in \vi(B)\}.
\]
Further, let ${\mathcal H}$ be the image of the mapping $S$ taken as
a multiset, i.e., 
\[
{\mathcal H}=\{S(h)\colon h\in H\}.
\]
Thus, $S$ is a bijection between $H$ and $\mathcal H$.
By saying that a mapping $S$ is a bijection between a set and
a multiset we mean that for any $\bar h\in{\mathcal H}$, the number of
$h\in H$ that map to $\bar h$ is equal to the multiplicity of $\bar h$.
Note that we cannot use some common properties of set bijections;
for instance we have to avoid using the inverse mapping $S^{-1}$.

Additionally, let
$\sigma$ be the induced bijection of $2^H$ and $2^{\mathcal H}$ 
defined by $\sigma(G)=\{S(h)\colon h\in G\}$, for $G\subseteq H$. 

Now, consider the triple $(\bfaceq,\leq,{\mathcal H})$, where $\leq$ is an
arbitrary linear extension of $\leq_1$ (such an extension exists since
$(H,\vi)$ is acyclic and $\leq_1$ therefore antisymmetric). This is a
concrete LP-type problem: The only thing to check is the existence of
a minimal element of every nonempty intersection $\bigcap{\mathcal G}$
($\mathcal G \subseteq \mathcal H$), which is guaranteed by the linearity 
of $\leq$ (remember from Definition \ref{def:concrete_LPtype} that
$\bigcap{\mathcal G}:=\bigcap_{G\in\mathcal{G}}G$).

It remains to prove basis-equivalence, which we do with the following two
lemmas.

\begin{lem}\label{lem:equ1}
If $B$ is an inclusion-minimal subset of $G$ with $\vi(B)=\vi(G)$ in $(H,\vi)$ 
(that is, $B$ is a basis of $G$), then $\min(\bigcap\sigma(B))=
\min(\bigcap\sigma(G))$ in $(\bfaceq,\leq,{\mathcal H})$.
\end{lem}

\pf It is clear that $[B]\in \bigcap\sigma(G)$. Therefore, showing 
that there is no other basis in $\bigcap\sigma(G)$ that is locally 
smaller than $[B]$ proves the lemma, because then 
$\min(\bigcap\sigma(G))=[B]=\min(\bigcap\sigma(B))$ (the second
equality holds since $B$ is a basis of $B$; just replace $G$ by $B$ in
the following proof). Assume for contradiction that a $C$ with
$[C]\neq [B]$, $[C]\in\bigcap\sigma(G)$ and $C\leq_0[B]$ exists.
By $[C]\in\bigcap\sigma(G)$ we have $G\cap\vi(C)=\emptyset$, which is
equivalent to
\[
(G\cup C)\cap\vi(C)=\emptyset,
\]
and by $C\leq_0[B]$ we have $C\cap\vi(B)=\emptyset$ which is equivalent
to (because $B$ is a basis of $G$)
\[
(G\cup C)\cap\vi(B)=\emptyset.
\]
Applying locality in $(H,\vi)$ to these two equations tells us that 
$\vi(C)=\vi(B)$, a contradiction to $[C]\neq [B]$.
\qed

\begin{lem}\label{lem:equ2}
If $\sigma(B)$ is an inclusion-minimal submultiset of $\sigma(G)$ with 
$\min(\bigcap\sigma(B))=\min(\bigcap\sigma(G))$ in 
$(\bfaceq,\leq,{\mathcal H})$ (that is, $\sigma(B)$ is a basis of 
$\sigma(G)$), then $\vi(B)=\vi(G)$ in $(H,\vi)$.
\end{lem}

\pf Let $A$ be a basis of $B$, so $\vi(A)=\vi(B)$. Note that 
$[A]\in\bigcap\sigma(B)$. Let $[C]=\min(\bigcap\sigma(B))$, 
thus $B\cap\vi(C)=\emptyset$ and therefore also $A\cap\vi(C)=\emptyset$. 
This means that $[A]\leq_0 [C]$ from which we conclude that $[A]=[C]$. 
From $\min(\bigcap\sigma(G))=[C]$ we get $G\cap\vi(C)=\emptyset$ which 
is equivalent to
\[
G\cap\vi(B)=\emptyset.
\]
As $\sigma(B)\subseteq\sigma(G)$ if and only if $B\subseteq G$, we can apply 
locality and derive $\vi(B)=\vi(G)$ as needed.
\qed

Lemmas \ref{lem:equ1} and \ref{lem:equ2} prove that $(H,\vi)$ and
$(\bfaceq,\leq,{\mathcal H})$ are basis-equivalent, in the sense
that $B$ is a basis of $G$ in $(H,\vi)$ if and only if $\sigma(B)$
is a basis of $\sigma(G)$ in $(\bfaceq,\leq,{\mathcal H})$:
Starting with a basis $B$ of $G$ in $(H,\vi)$, Lemma \ref{lem:equ1}
yields $\min(\bigcap\sigma(B))=\min(\bigcap\sigma(G))$. This
$\sigma(B)$ is inclusion-minimal w.r.t. $\sigma(G)$, since otherwise
Lemma \ref{lem:equ2} would yield a contradiction to the 
inclusion-minimality of $B$ w.r.t. $G$ (where we again use that 
$\sigma(B)\subseteq\sigma(G)$ if and only if $B\subseteq G$). The 
reasoning for inclusion-minimality in the other direction is analogous. 
This concludes the proof of Proposition \ref{prop:acyclicVS->concrete}.
\qed

Propositions \ref{prop:abstract->acyclicVS} and \ref{prop:acyclicVS->concrete},
together with the fact that every concrete LP-type problem can
be transformed into an abstract one (as described below 
Definition \ref{def:concrete_basis}), yield Theorem \ref{thm:main}.

\subsection{Examples}\label{sec:examples} 
Here we present some particular abstract LP-type problems and we 
demonstrate the construction (via acyclic violator spaces) of their 
concrete representations.

Let $a$, $b$, $c$ and $d$ be the vertices of a unit square (in the
counterclockwise order); let $H=\{a,b,c,d\}$. For $G\subseteq H$ let
$w(G)$ be the radius of the smallest circle enclosing all the points
of $G$ (for $G=\emptyset$ put $w(G)=-\infty$). The corresponding
acyclic violator space is described by the following table:
\begin{center}
\tabcolsep0.4cm
\begin{tabular}{|c||c|c|c|c|c|c|c|c|}\hline
$G$ & $\emptyset$ & $a$ & $b$ & $c$ & $d$ & $ab$ & $ac$ & $ad$ \\ \hline
$\vi(G)$ & $abcd$ & $bcd$ & $acd$ & $abd$ & $abc$ & $cd$ & $\emptyset$ & $bc$ \\ \hline \hline
$G$ & $bc$ & $bd$ & $cd$ & $abc$ & $abd$ & $acd$ & $bcd$ & $abcd$  \\ \hline
$\vi(G)$ & $ad$ & $\emptyset$ & $ab$ & $\emptyset$ & $\emptyset$ & $\emptyset$ & $\emptyset$ & $\emptyset$ \\ \hline
\end{tabular}
\end{center}
The bases are $\emptyset$, $a$, $b$, $c$, $d$, $ab$, $ac$, $ad$, $bc$,
$bd$, $cd$; the only equivalent pair is $ac\sim bd$. There is no
inconvenience concerning differences between $\leq_0$ on sets and
equivalence classes and $\leq_1$; the ordering $\leq_1$ is given by the
Hasse diagram in Figure \ref{fig:hasse}.

\figeps{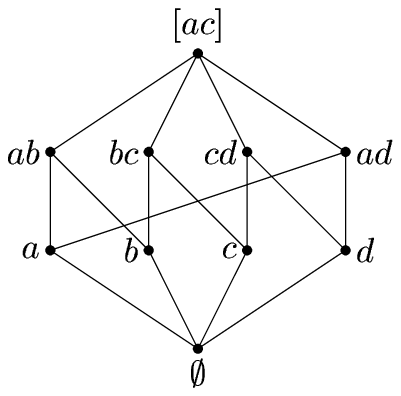}{4}{Hasse diagram from smallest enclosing circle of the
  vertices of a square.}

As the linear extension $\leq$ of $\leq_1$ we may choose 
$\emptyset<a<b<c<d<ab<bc<cd<ad<[ac]$. 
Finally, the concrete representation $S$ is as follows:
\begin{center}
\tabcolsep0.4cm
\begin{tabular}{|c||c|c|c|c|}\hline
$h$ & $a$ & $b$ & $c$ & $d$ \\ \hline
$S(h)$ & $a,ab,ad,[ac]$ & $b,ab,bc,[ac]$ & $c,bc,cd,[ac]$ 
& $d,cd,ad,[ac]$ \\ \hline
\end{tabular}
\end{center}
In the geometric view that we have mentioned earlier, $S(a)$ corresponds to
the set of all ``canonical'' (i.e., basic) balls that contain the point 
$a$ (inside or on the boundary). The same holds for the other points.

As the other example, consider the following LP problem in the positive 
orthant (rotated by 45 degrees for convenience). Beside the restriction
to the positive orthant, the constraints are the four halfplanes depicted in 
Figure \ref{fig:lp_example}. The optimization direction is given by the arrow.

\figeps{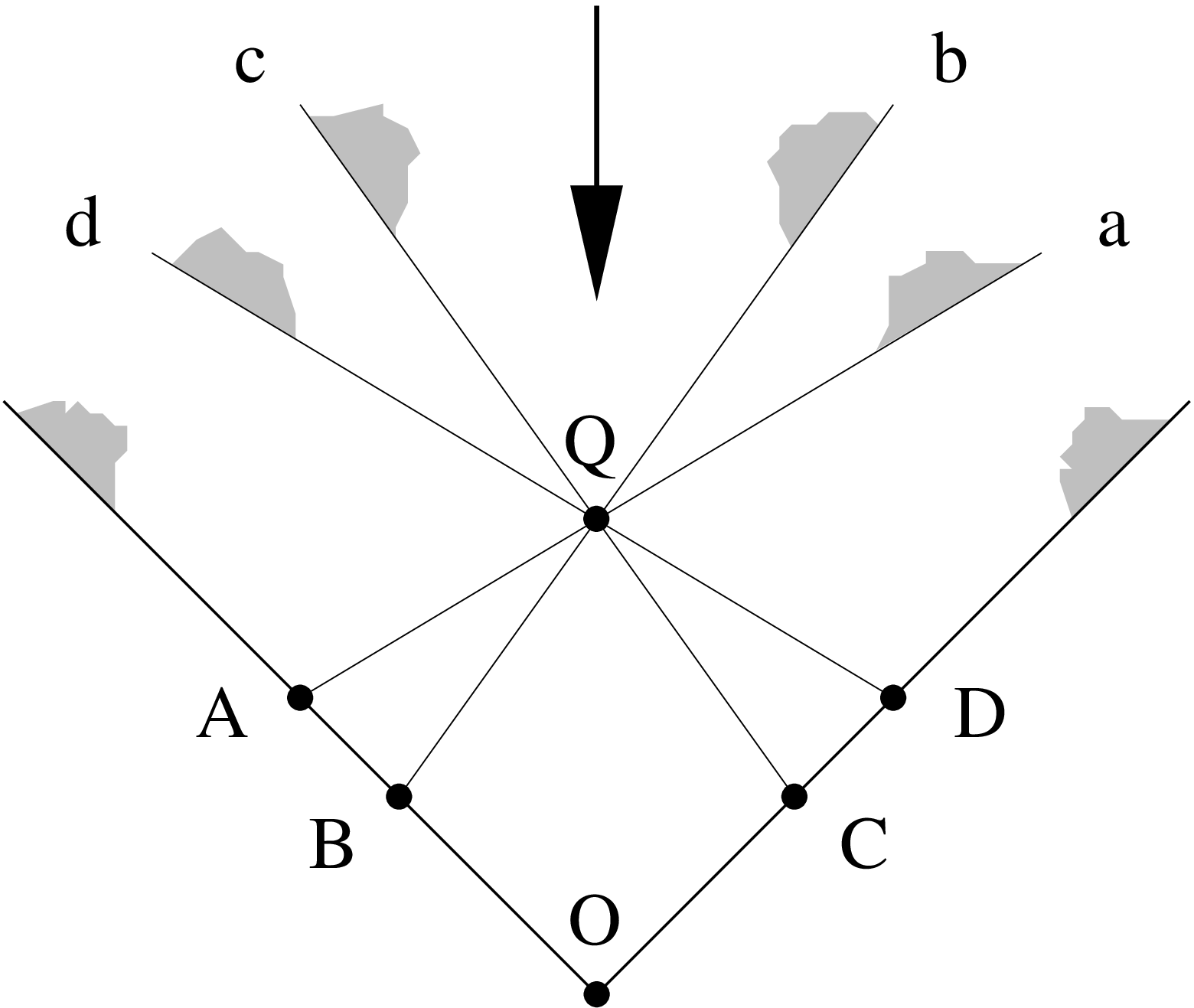}{6}{Illustration example -- linear programming.}

Here the violator space bases are $\emptyset$, $a$, $b$, $c$, $d$, $ac$, $ad$, $bc$,
$bd$; the equivalence classes are $O=\emptyset$, $A=\{a\}$, $B=\{b\}$,
$C=\{c\}$, $D=\{d\}$ and $Q=\{ac\}\sim\{ad\}\sim\{bc\}\sim\{bd\}$.
Note that the equivalence classes correspond to the points in the plane.
We have $O\leq_1 B\leq_1 A\leq_1 Q$ and $O\leq_1 C\leq_1 D\leq_1 Q$; we
choose $\leq$ to be $O<B<A<C<D<Q$. The concrete representation is 
\begin{center}
\tabcolsep0.4cm
\begin{tabular}{|c||c|c|c|c|}\hline
$h$ & $a$ & $b$ & $c$ & $d$ \\ \hline
$S(h)$ & $A,Q$ & $A,B,Q$ & $C,D,Q$ & $D,Q$ \\ \hline
\end{tabular}.
\end{center}
Here we may interpret $S(a)$ as the set of all ``canonical'' points
lying in the halfplane $a$.

To see why we allow $\mathcal H$ in the definition of a concrete
LP-type problem to be a multiset, consider the abstract LP-type
problem with $H=\{a,b\}$ and $w(G)=0$ for every $G\subseteq H$.
The only basis is $\emptyset$ and it is not violated by any
$h\in H$. Thus we have $S(a)=S(b)=\{[\emptyset]\}$. If we do not allow
$\mathcal H$ to be a multiset, we have ${\mathcal H}=\{S(a),S(b)\} =
\bigl\{ \{[\emptyset]\} \bigr\}$ with only one constraint; it seems
improper to define this to be basis-equivalent to $H$.
We could alter the construction of $\mathcal H$ and get 
$S(a)=\{0\}$, $S(b)=\{0,1\}$,
which does represent the original abstract LP-type problem with
$\mathcal H$ being a set; however, we believe that our definition
catches the structure in a more straightforward way, although it may
seem unusual at first glance.

\section{Clarkson's Algorithms}\label{sec:clarkson}
We show that Clarkson's randomized reduction scheme, originally
developed for linear programs with many constraints and few variables,
actually works for general (possibly cyclic) violator spaces.  The two
algorithms of Clarkson involved in the reduction have been analyzed
for LP and LP-type problems before
\cite{c-lvali-95,gw-lpraf-96,cm-ltdao-96}; the analysis we give below
is almost identical on the abstract level. Our new contribution is
that the combinatorial properties underlying Clarkson's algorithms
also hold for violator spaces.

We start off by deriving these combinatorial properties; the 
analysis of Clarkson's reduction scheme is included for completeness.

\subsection{Violator spaces revisited}
We recall that an abstract LP-type problem is of the form $(H,w,W,\leq)$.
In this subsection we will view a violator space as 
an ``LP-type problem without the order $\leq$'', i.e., we will 
only care whether two subsets $F$ and $G$, $F\subseteq G\subseteq H$, have
the same value (and therefore the same violators, see Lemma
\ref{lem:separbasis}), but not how they compare under the order $\leq$.
It turns out that the order is irrelevant for Clarkson's algorithms. 

Even without an order, we can talk about monotonicity in violator 
spaces:
\begin{lem}
Any violator space $(H,\vi)$ satisfies 

\begin{tabular}{ll} Monotonicity: & $\vi(F)=\vi(G)$ implies 
$\vi(E)=\vi(F)=\vi(G)$,\\
& for all sets $F\subseteq E\subseteq G\subseteq H$.
\end{tabular}
\label{lem:monotonicity}
\end{lem}

\pf Assume $\vi(E)\neq \vi(F),\vi(G)$. Then locality yields
$\emptyset \neq E\cap \vi(F) = E\cap \vi(G)$ which contradicts
consistency. \qed

Recall Definition \ref{def:vbasis}: A basis is a set $B$ satisfying
$B\cap \vi(F)\neq\emptyset$ for all proper subsets $F$ of $B$. A basis
of $G$ is an inclusion-minimal subset of $G$ with the same violators.
This can be used to prove the following observation, well-known to
hold for LP-type problems \cite{gw-lpraf-96}.

\begin{obs}\label{obs:extreme}
Let $(H,\vi)$ be a violator space. For $R\subseteq H$ and all $h\in H$, we have
\begin{enumerate}
\item[(i)] $\vi(R)\neq \vi(R\cup\{h\})$ if and only if $h\in \vi(R)$, and
\item[(ii)] $\vi(R)\neq \vi(R\setminus\{h\})$ if and only if $h$ is contained
in every basis of $R$. 
\end{enumerate}
An element $h$ such that (ii) holds is called \emph{extreme} in $R$.
\end{obs}

\pf (i) If $h\notin\vi(R)$, we get $\vi(R)=\vi(R\cup\{h\})$ by 
Lemma \ref{lem:conseqloc}. If $h\in\vi(R)$, then $\vi(R)\neq \vi(R\cup\{h\})$
is a consequence of consistency applied to $G=R\cup\{h\}$. (ii) if
$\vi(R)=\vi(R\setminus\{h\})$, there is a basis $B$ of $R\setminus\{h\}$,
 and this basis is also a basis of $R$ not containing $h$. Conversely,
if there is some basis $B$ of $R$ not containing $h$, then 
$\vi(R)=\vi(R\setminus\{h\})$ follows from monotonicity.
\qed

We are particularly interested in violator spaces with small bases.

\begin{defn}
Let $(H,\vi)$ be a violator space. The size of a largest basis is called
the \emph{combinatorial dimension} $\delta=\delta(H,\vi)$ of $(H,\vi)$.
\end{defn}

Observation \ref{obs:extreme} implies that in a violator space of
combinatorial dimension $\delta$, every set has at most $\delta$
extreme elements. This in turn yields a bound for the \emph{expected} 
number of violators of a random subset of constraints, using the 
\emph{sampling lemma} \cite{GWSampl01}.

\begin{lem}{\rm\cite{GWSampl01}} 
Consider a triple $(H,w,W)$, where $w$ is a function mapping subsets of 
the set $H$ to the set $W$ (not necessarily ordered). 
For $R\subseteq H$, we define
\begin{eqnarray*}
V(R) &:=& \{h\in H\setminus R: w(R)\neq w(R\cup\{h\}), \\
X(R) &:=& \{h\in R: w(R)\neq w(R\setminus\{h\}).
\end{eqnarray*}
For $0\leq r\leq |H|$, let $v_r$ be the expected value of $|V(R)|$,
for $R$ chosen uniformly at random among all subsets of $H$ with
$r$ elements. $x_r$ is defined similarly as the expected value
of $|X(R)|$. Then for $0\leq r<n$, the following equality holds.
\[\frac{v_r}{n-r}=\frac{x_{r+1}}{r+1}.\]
\end{lem}

To apply this in our situation, we fix a set $W\subseteq H$, and we
define $w(R)=\vi(W\cup R)$. Since then $|X(R)|\leq\delta$ for all
$R$, the following Corollary is obtained.

\begin{cor}\label{cor:sampling}
Let $(H,\vi)$ be a violator space of combinatorial dimension $\delta$
and $W\subseteq H$ some fixed set. Let $v_r$ be the expected number of 
violators of the set $W\cup R$, where $R\subseteq H$ is a random subset
of size $r< n=|H|$. Then
\[
v_r \leq \delta \frac{n-r}{r+1}.
\]
\end{cor}

\subsection{The Trivial Algorithm}
Given a violator space $(H,\vi)$ of combinatorial dimension $\delta$,
the goal is to find a basis of $H$. For this, we assume availability
of the following primitive.

\begin{primitive}\label{prim:viol}
Given $G\subseteq H$ and $h\in H\setminus G$, decide whether $h\in \vi(G)$.
\end{primitive}

Given this primitive, the problem can be solved in a brute-force manner 
by going through all sets of size $\leq \delta$, testing each of them for 
being a basis of $H$. More generally, $B\subseteq G$ is a basis of $G$ if 
and only if
\[
\begin{array}{rcll}
h&\in& \vi(B\setminus\{h\}), \quad & \forall h\in B, \\
h&\notin& \vi(B), & \forall h\in G\setminus B. 
\end{array}
\]
Consequently, the number of times the primitive needs to be invoked 
in order to find a basis of $H$ is at most 
\[
n \sum_{i=0}^{\delta}{n\choose i} = O(n^{\delta+1}).
\]
The next two subsections show that this can be substantially improved.

\subsection{Clarkson's First Algorithm}
Fix a violator space $(H,\vi)$ of combinatorial dimension $\delta$,
implicitly specified through Primitive \ref{prim:viol}. Clarkson's
first algorithm calls Clarkson's second algorithm (\keyw{Basis2})
as a subroutine. Given $G\subseteq H$, both algorithms compute a basis
$B$ of $G$.

\begin{tabbing}
\quad \=\quad \=\quad \=\quad \=\quad                \kill
\keyw{Basis1}$(G)$:                               \\
\> (* computes a basis $B$ of $G$ *)                 \\
\> \keyw{IF} $|G|\leq 9\delta^2$ \keyw{THEN}         \\
\>\> \keyw{RETURN} \keyw{Basis2}$(G)$             \\
\> \keyw{ELSE}                                       \\
\>\> $r:=\lfloor\delta\sqrt{|G|}\rfloor$             \\
\>\> $W:=\emptyset$                                  \\
\>\> \keyw{REPEAT}                                   \\
\>\>\> choose $R$ to be a random $r$-element subset 
of $G$, $R\in{G\choose r}$                           \\
\>\>\> $C:=\keyw{Basis2}(W\cup R)$                \\
\>\>\> $\vi:=\{h\in G\setminus C\colon h\in \vi(C)\}$\\
\>\>\> \keyw{IF} $|\vi|\leq 2\sqrt{|G|}$ \keyw{THEN} \\
\>\>\>\> $W:=W\cup \vi$                              \\
\>\>\> \keyw{END}                                    \\
\>\> \keyw{UNTIL} $\vi=\emptyset$                    \\
\>\> \keyw{RETURN} $C$                               \\
\> \keyw{END}                                        \\
\end{tabbing}

Assuming \keyw{Basis2} is correct, this algorithm is correct as well: 
if $B$ is a basis of $W\cup R\subseteq G$ that in addition has no violators 
in $G$, $B$ is a basis of $G$. Moreover, the algorithm augments the working 
set $W$ at most $\delta$ times, which is guaranteed by the following 
observation.

\begin{obs}\label{obs:basis}
If $C\subseteq G$ and $G\cap \vi(C)\neq \emptyset$, then $G\cap \vi(C)$
contains at least one element from every basis of $G$.
\end{obs}

\pf Let $B$ be a basis of $G$. Assuming 
\[
\emptyset = B\cap G\cap \vi(C) = B \cap \vi(C),
\]
consistency yields $C\cap \vi(C)=\emptyset$, implying 
$(B\cup C)\cap \vi(C)=\emptyset$. From locality and monotonicity (Lemma 
\ref{lem:monotonicity}), we get
\[
\vi(C)=\vi(B\cup C)=\vi(G),
\]
meaning that $G\cap \vi(G)= G\cap \vi(C)=\emptyset$, a contradiction.
\qed

It is also clear that \keyw{Basis2} is called only
with sets of size at most $3\delta\sqrt{|G|}$. Finally, the expected
number of iterations through the \keyw{REPEAT} loop is bounded by
$2\delta$: by Corollary \ref{cor:sampling} (applied to $(G,\vi|_G)$) and the 
Markov inequality, the expected number of calls to \keyw{Basis2} 
before we next augment $W$ is bounded by $2$.

\begin{lem}
Algorithm \keyw{Basis1} computes a basis of $G$ with an expected number 
of at most $2\delta|G|$ calls to Primitive \ref{prim:viol}, and an expected 
number of at most $2\delta$ calls to \keyw{Basis2}, with sets of size 
at most $3\delta\sqrt{|G|}$. 
\end{lem}

\subsection{Clarkson's Second Algorithm}
This algorithm calls the trivial algorithm as a subroutine.
Instead of adding violated constraints to a working set,
it gives them larger probability of being selected in
further iterations. Technically, this is done by maintaining
$G$ as a multiset, where $\mu(h)$ denotes the multiplicity
of $h$ (we set $\mu(F):=\sum_{h\in F}\mu(h)$). Sampling from 
$G$ is done as before, imagining that $G$ contains $\mu(h)$ 
copies of the element $h$.

\begin{tabbing}
\quad \=\quad \=\quad \=\quad \=\quad                      \kill
\keyw{Basis2}$(G)$:                                     \\
\> (* computes a basis $B$ of $G$ *)                       \\
\> \keyw{IF} $|G|\leq 6\delta^2$ \keyw{THEN}               \\
\>\> \keyw{RETURN} \keyw{Trivial}$(G)$                     \\
\> \keyw{ELSE}                                             \\
\>\> $r:=6\delta^2$                                        \\
\>\> \keyw{REPEAT}                                         \\
\>\>\> choose random $R\in{G\choose r}$                   \\
\>\>\> $C:=\keyw{Trivial}(R)$                              \\
\>\>\> $\vi:=\{h\in G\setminus C\colon h\in \vi(C)\}$      \\
\>\>\> \keyw{IF} $\mu(\vi)\leq \mu(G)/3\delta$ \keyw{THEN} \\
\>\>\>\> $\mu(h) := 2\mu(h), \quad h\in \vi$               \\
\>\>\> \keyw{END}                                          \\
\>\> \keyw{UNTIL} $\vi=\emptyset$                          \\
\>\> \keyw{RETURN} $C$                                     \\
\> \keyw{END}                                              \\
\end{tabbing}

Invoking Corollary \ref{cor:sampling} again (which also applies to
multisets as we use them), we see that the expected number of
calls to \keyw{Trivial} before we next reweight elements
(a \emph{successful} iteration), is bounded by $2$. It remains 
to bound the number of successful iterations.

\begin{lem}
Let $k$ be a positive integer. After $k\delta$ successful iterations, 
we have \[
2^k \leq \mu(B)\leq |G|e^{k/3},
\]
for every basis $B$ of $G$. In particular, $k< 3\ln |G|$.
\end{lem}

\pf Every successful iteration multiplies the total weight
of elements in $G$ by at most $(1+1/3\delta)$, which gives the upper 
bound (not only for $\mu(B)$ but actually for $\mu(G)$). For the
lower bound, we use Observation \ref{obs:basis} again to argue
that each successful iteration doubles the weight of some element in $B$, 
meaning that after $k\delta$ iterations, one element has been doubled at 
least $k$ times. Because the lower bound exceeds the upper bound for 
$k\geq 3\ln |G|$, the bound on $k$ follows.
\qed

Summarizing, we get the following lemma.

\begin{lem}
Algorithm \keyw{Basis2} computes a basis of $G$ with an expected number 
of at most $6\delta |G|\ln|G|$ calls to Primitive \ref{prim:viol}, and 
expected number of at most $6\delta\ln |G|$ calls to \keyw{Trivial}, 
with sets of size $6\delta^2$.
\end{lem}

\subsection{Combining the Algorithms}
\begin{thm}\label{thm:runtime}
Using a combination of the above two algorithms, a basis of $H$ in a violator 
space $(H,\vi)$ can be found calling Primitive \ref{prim:viol} expected
\[
O\left(\delta n + \delta^{O(\delta)}\right)
\] 
many times.
\end{thm}

\pf Using the above bound for the trivial algorithm, \keyw{Basis2}
can be implemented to require an expected number of at most
\[O\left(\delta\log |G|(|G| + \delta^{O(\delta)})\right)\]
calls to the primitive. Applying this as a subroutine in
\keyw{Basis1}$(H)$ with $|H|=n$, $|G|$ is bounded by
$3\delta\sqrt{n}$, and we get an overall expected complexity of
\[
O\left(\delta n + \delta^2 (\log n(\delta \sqrt{n} + 
\delta^{O(\delta)}))\right)
\]
in terms of the number of calls to Primitive \ref{prim:viol}. The
terms $\delta^2\log n~ \delta\sqrt{n}$ and $\delta^2\log n~ \delta^{O(\delta)}$ 
are asymptotically
dominated by either $\delta n$ or $\delta^{O(\delta)}$, and we
get the simplified bound of
$O\left(\delta n + \delta^{O(\delta)}\right)$.
\qed

\section{Grid USO as Models for Violator Spaces}\label{grid_uso}
We show in this section that the problem of finding the sink
in a $\delta$-dimensional \emph{grid unique sink orientation}
\cite{grid_uso} can be reduced to the problem of finding the (unique) 
basis of a violator space of combinatorial dimension $\delta$.

Unique sink orientations of grids arise from various problems, 
including linear programming over products of simplices and generalized
linear complementarity problems (GLCP) over $\P$-matrices \cite{grid_uso}.
The GLCP has been introduced by Cottle and Dantzig \cite{cd-glcp-70}
as a generalization of the well known LCP \cite{lcp}.
There are also applications in game theory; for instance
\cite{sweden1,sweden2,ssg_pglcp} show how parity, mean-payoff, 
and simple stochastic games are related to grid USO.

\subsection{Grid USO}
Fix a partition
\[
\Pi=(\Pi_1,\ldots,\Pi_{\delta})
\]
of the set $H:=\{1,\ldots,n\}$ into $\delta$ nonempty subsets,
where we refer to $\Pi_i$ as the \emph{block $i$}. A subset
$J\subseteq H$ is called a \emph{vertex} if $|J\cap \Pi_i|=1$
for all $i$. The vertices naturally correspond to the Cartesian
product of the $\Pi_i$. Let $\V$ be the set of all vertices. 

In the following definition, we introduce the \emph{grid spanned} 
by subsets $\Pi'_i$ whose union is $G\subseteq H$. The vertex set 
of this grid contains all vertices $J\subseteq G$ ($J\in \V$), 
with two vertices
being adjacent whenever they differ in exactly two elements.

\begin{defn}
The $\delta$-dimensional \emph{grid} spanned by $G \subseteq H$
is the undirected graph $\G(G)=(\V(G),\E(G))$, with
\[
\V(G):=\{J\in \V\colon J\subseteq G\},\quad
\E(G):=\{\{J,J'\}\subseteq\V(G)\colon |J\oplus J'|=2\}.
\]
Here, $\oplus$ is the symmetric difference of sets.
\end{defn}

$\V(G)$ is in one-to-one correspondence with the Cartesian product 
\[
\prod_{i=1}^{\delta}G_i, \quad G_i:=G\cap \Pi_i,
\]
and the edges in $\E(G)$ connect vertices in $\V(G)$ whose
corresponding tuples differ in exactly one coordinate.
See Figure \ref{fig:grid} left for an example of a grid.

\figeps{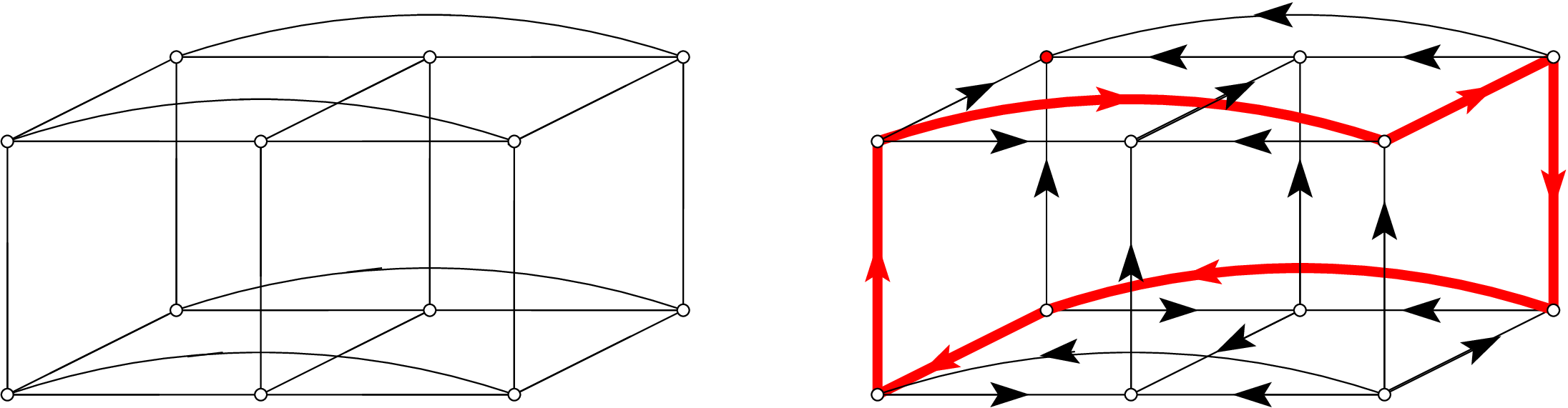}{13}{A 3-dimensional grid $\G(H)$ with $H=\{1,\ldots,7\}$ 
where $\Pi=(\{1,2,3\},\{4,5\},\{6,7\})$ and a USO of it.}

Note that $\G(G)$ is the empty graph whenever 
$G_i=G\cap \Pi_i=\emptyset$ for some $i$. We say that such a 
$G$ is \emph{not $\Pi$-valid}, and it is \emph{$\Pi$-valid}
otherwise.

A \emph{subgrid} of $\G(G)$ is any graph of the form $\G(G')$,
for $G'\subseteq G$.

\begin{defn}
An orientation $\psi$ of the graph $\G:=\G(H)$ is called a
\emph{unique sink orientation} (USO) if all nonempty subgrids of 
$\G$ have unique sinks w.r.t.\ $\psi$. 
\end{defn}

We are interested in finding the sink in a USO of $\G$ as fast as possible,
since the sink corresponds to the solution of the underlying problem
(the $\P$-matrix GLCP, for example). Our measure of complexity
will be the expected number of \emph{edge evaluations}, see \cite{grid_uso}.
An edge evaluation returns the orientation of the considered edge and
can typically be implemented to run in polynomial time (depending on
the underlying problem). In the remainder of this paper, we derive 
the following theorem.

\begin{thm}\label{thm:uso_algo}
The sink of a unique sink grid orientation can be found by evaluating
expected $O\left(\delta n + \delta^{O(\delta)}\right)$ edges.
\end{thm}

Note that a USO $\psi$ can be cyclic (see the thick edges in
Figure \ref{fig:grid} right).
If $\psi$ induces the directed edge $(J,J')$, we also write 
$J\stackrel{\psi}{\rightarrow}J'$. Any USO can be specified by associating 
each vertex $J$ with its outgoing edges. Given $J$ and $j\in H\setminus J$, 
we define $J\rhd j$ to be the unique vertex $J'\subseteq J\cup\{j\}$
that is different from $J$, and we call $J'$ the \emph{neighbor} of
$J$ \emph{in direction} $j$. Note that $J$ is a neighbor of $J'$ in
some direction different from $j$.

\begin{defn}
Given an orientation $\psi$ of $\G$, the 
function $s_{\psi}:\V\rightarrow 2^{H}$, defined by
\begin{equation}
s_{\psi}(J) := \{j\in H\setminus J\colon J\stackrel{\psi}{\rightarrow} J\rhd j\},
\end{equation}
is called the \emph{outmap} of $\psi$.
\end{defn}
By this definition, any sink w.r.t.\ $\psi$ has empty outmap value.

\subsection{Reduction to Violator Spaces}
Let us fix a unique sink orientation $\psi$ of $\G$. 
Given a $\Pi$-valid subset $G\subseteq H$, we define $\sink(G)\in
\V(G)$ to be the unique sink vertex in $\G(G)$. For a subset $G$ that
is not $\Pi$-valid, let
\[
\bar{G}:=\bigcup_{i\colon\, G_i=\emptyset}\Pi_i.
\]
Thus $\bar{G}$ is the set of elements occurring in blocks of $\Pi$ 
disjoint from $G$. 

\begin{defn}\label{def:gridtoviol}
For $G\subseteq H$, define
\[
\vi(G)=
\left\{
\begin{array}{ll}
s_{\psi}(\sink(G)), &\quad\textit{if $G$ is $\Pi$-valid}\\
\bar{G}, &\quad\textit{if $G$ is not $\Pi$-valid.}
\end{array}
\right.
\]
\end{defn}

\begin{thm}
The pair $(H,\vi)$ from Definition \ref{def:gridtoviol}
is a violator space of combinatorial dimension $\delta$.
Moreover, for all $\Pi$-valid $G\subseteq H$, the unique sink of the subgrid 
$\mathcal{G}(G)$ corresponds to the unique basis of $G$ in $(H,\vi)$.
\end{thm}

\pf For every $G \subseteq H$, consistency holds by
definition of $\sink(G), s_{\psi}(J)$ and $\bar{G}$.
In order to prove locality for $F\subseteq G\subseteq H$, 
we look at three different cases.

\paragraph*{$\mathbf{G}$ is not $\mathbf{\Pi}$-valid.} Then, $F\subseteq G$
is not $\Pi$-valid either. The condition
$\emptyset=G\cap \vi(F)=G\cap\bar{F}$ means that
$F$ is disjoint from the same blocks as $G$. This 
implies $\bar{G}=\bar{F}$, hence $\vi(G)=\vi(F)$.

\paragraph*{$\mathbf{G}$ and $\mathbf{F}$ are both 
$\mathbf{\Pi}$-valid.} Then $\G(F)$ is a nonempty subgrid of $\G(G)$, and 
$G \cap \vi(F)=\emptyset$ means that the sink of $\G(F)$ has no outgoing 
edges into $\G(G)$. Thus the unique sink of $\G(F)$ is also a sink of 
$\G(G)$ and therefore the unique one. This means that 
$\sink(G)=\sink(F)$, from which $\vi(G)=\vi(F)$ follows.

\paragraph*{$\mathbf{G}$ is $\mathbf{\Pi}$-valid, 
$\mathbf{F}$ is not  $\mathbf{\Pi}$-valid.} Then the condition 
$G \cap \vi(F)=\emptyset$ can never be satisfied since 
$\vi(F)=\bar{F}$ contains at least one full block $\Pi_i$,
and $G_i=G\cap\Pi_i\neq \emptyset$.

Next we prove that a largest basis in $(H,\vi)$ has at most $\delta$
elements. For this, let $G\subseteq H$ be a set of size larger than 
$\delta$. If $G$ is $\Pi$-valid, we have 
\[\vi(G):=s_{\psi}(\sink(G))=s_{\psi}(\sink(\sink(G)))=:\vi(\sink(G)),\] 
since $J=\sink(J)$ for any vertex $J$. This means that $G$ has a subset
of size $\delta$ with the same violators, so $G$ is not a basis.

If $G$ is not $\Pi$-valid, we consider some subset $B$ that contains 
exactly one element from every block intersected by $G$. By definition,
we have $\bar{G}=\bar{B}$ and $\vi(G)=\vi(B)$. Since $B$ has less than
$\delta$ elements, $G$ cannot be a basis in this case, either.

It remains to prove that for $G$ being $\Pi$-valid, the vertex
$\sink(G)$ is the unique basis of $G$ in $(H,\vi)$. We have 
already shown that $\vi(G)=\vi(\sink(G))$ must hold in this 
case. Moreover, $\vi(\sink(G))$ contains no full block $\Pi_i$.
On the other hand, any proper subset $F$ of $\sink(G)$ is not
$\Pi$-valid, so its violator set \emph{does} contain at least
one full block. It follows that $V(F)\neq V(\sink(G))$, so
$\sink(G)$ is a basis of $G$. The argument is complete when
we can prove that no other vertex $J\subseteq G$ is a basis of 
$G$. Indeed, such a vertex $J$ is not a sink in $\G(G)$, meaning 
that $G\cap \vi(J)\neq \emptyset$. This implies $\vi(J)\neq \vi(G)$.
\qed

Note that the global sink of the grid USO corresponds to the unique
$\delta$-element (and $\Pi$-valid) set $B$ with $\vi(B)=\emptyset$.
This is exactly the set output by the call \texttt{Basis1}$(H)$ of
Clarkson's algorithms, when we apply it to the violator space
constructed in Definition \ref{def:gridtoviol}.

Primitive \ref{prim:viol} corresponds to one edge evaluation in the USO 
setting. With Theorem \ref{thm:runtime}, we therefore have proved
Theorem \ref{thm:uso_algo}. For small $\delta$, the running time given
in the theorem is faster than the one from the \emph{Product Algorithm} 
\cite{grid_uso} which needs expected $O(\delta!n+H_n^{\delta})$ edge
evaluations, where $H_n$ is the $n$-th harmonic number.

\section{Conclusions}
We introduced violator spaces as a new framework for optimization 
problems and showed that acyclic violator spaces are
equivalent to abstract and concrete LP-type problems.
It turned out that the explicit ordering inherent to 
LP-type problems is not necessary in order to capture 
the structure of the underlying optimization 
problem. Violator spaces are more general than LP-type problems,
yet Clarkson's algorithms still work on them. 

The Sharir-Welzl algorithm is also applicable for violator spaces 
in a straightforward way. However, the most obvious translation
of this algorithm to the setting of violator spaces
is not even guaranteed to finish, since for a general violator space 
it may run in a cycle and the subexponential analysis thus breaks down.

We have seen that unique sink orientations are models for possibly cyclic
violator spaces, and with Clarkson's algorithms we therefore
have a fast scheme to solve fixed dimensional USO problems like
the generalized linear complementarity problem with a $\P$-matrix. 
The GLCP with a $\P$-matrix has in general a cyclic structure and
therefore gives rise to a cyclic USO. A violator space obtained from 
a cyclic USO is again cyclic. It is interesting that there are no cycles 
in a 2-dimensional grid USO \cite{grid_uso}. Whether the same is true 
for violator spaces of combinatorial dimension 2 is an open question.

\section*{Acknowledgment} 
We thank an anonymous referee for useful comments.
The second author would like to thank Nina Amenta for discussions 
concerning LP-type problems, possibly already forgotten by her as 
they took place many years ago, but nevertheless helpful for reaching 
the results in this paper.

\end{document}